\documentclass[12pt]{article}
\usepackage{amsmath}
\usepackage{graphicx}
\usepackage{citesort}
\begin{document}

\setcounter{page}{0}
\thispagestyle{empty}

\begin{flushright}
{\small BARI-TH 436/2002}
\end{flushright}

\vspace*{2.5cm}

\begin{center}
{\large\bf Abelian chromomagnetic fields and confinement}
\end{center}

\vspace*{2cm}

\renewcommand{\thefootnote}{\fnsymbol{footnote}}

\begin{center}
{%\large
Paolo Cea$^{1,2,}$\protect\footnote{{\tt Paolo.Cea@ba.infn.it}} and
Leonardo Cosmai$^{1,}$\protect\footnote{{\tt Leonardo.Cosmai@ba.infn.it}} \\[0.5cm]
$^1${\em INFN - Sezione di Bari, I-70126 Bari, Italy}\\[0.3cm]
$^2${\em Physics Department, Univ. of Bari, I-70126 Bari, Italy}
}
\end{center}

\vspace*{0.5cm}

\begin{center}
{%\large
April, 2002}
\end{center}

\vspace*{1.0cm}

\renewcommand{\abstractname}{\normalsize Abstract}
\begin{abstract}
We study vacuum dynamics of SU(3) lattice gauge theory at finite
temperature using the lattice Schr\"odinger functional.
The SU(3)
vacuum is probed by means of an external constant Abelian
chromomagnetic field.  We find that by increasing the strength of
the applied external field the deconfinement temperature decreases
towards zero.  This means that strong enough Abelian
chromomagnetic fields destroy confinement of color. We 
discuss some consequences of this phenomenon on confinement and quark stars.
\end{abstract}

\vfill
\newpage

Understanding the mechanism of quark confinement is still a
central problem in high energy physics. According to a model
conjectured long time ago by G.~'t~Hooft~\cite{tHooft:1976eps} and
S.~Mandelstam~\cite{Mandelstam:1974pi} the confining vacuum
behaves as a coherent state of color magnetic monopoles, or,
equivalently, the vacuum resembles a magnetic (dual)
superconductor. Up to now there is  numerical 
evidence~\cite{Wosiek:1987kx,DiGiacomo:1990hc,Cea:1993sd,Singh:1993ma,Matsubara:1994nq,Bali:1995de,Cea:1995ed,Cea:1995zt,DiGiacomo:1996ca} in
favor of chromoelectric flux tubes in pure lattice gauge vacuum.
As well there have been extensive numerical 
studies~\cite{Stack:1992zp,Shiba:1995db,Arasaki:1997sm,Nakamura:1997sw,Chernodub:1997ps,Jersak:1999nv,DiGiacomo:1999fa,DiGiacomo:1999fb,Cea:1999zv,Hoelbling:2000su,Cea:2000zr,Carmona:2001ja} 
of monopole condensation.

To investigate vacuum structure of lattice gauge theories
we introduced~\cite{Cea:1997ff,Cea:1999gn} a gauge invariant
effective action, defined by means of the lattice Schr\"odinger
functional~\cite{Luscher:1992an,Luscher:1995vs}
\begin{equation}
\label{Zetalatt} {\mathcal{Z}}[U^{\mathrm{ext}}_k] = \int
{\mathcal{D}}U \; e^{-S_W} \,.
\end{equation}
$S_W$ is the 
Wilson action and the functional
integration is extended over links on a lattice
$L_s^3 \times L_t$  with 
hypertorus geometry  and satisfying the
constraints ($x_t$: temporal coordinate)
\begin{equation}
\label{coldwall}
U_k(x)|_{x_t=0} = U^{\mathrm{ext}}_k(\vec{x})
\,,\,\,\,\,\, (k=1,2,3) \,\,,
\end{equation}
$U^{\mathrm{ext}}_k(x)$ being the lattice version of
the external continuum gauge field
$\vec{A}^{\mathrm{ext}}(x)=
\vec{A}^{\mathrm{ext}}_a(x) \lambda_a/2$.
We also impose that links at the spatial boundaries are fixed
according to Eq.~(\ref{coldwall}). In the continuum this last
condition amounts to the requirement that fluctuations over the
background field vanish at infinity.

In terms of the above defined lattice Schr\"odinger functional the
lattice effective action for  external static background field
$\vec{A}^{\mathrm{ext}}(x)$ is given by:
\begin{equation}
\label{Gamma} \Gamma[\vec{A}^{\mathrm{ext}}] = -\frac{1}{L_t} \ln
\left\{
\frac{{\mathcal{Z}}[\vec{A}^{\mathrm{ext}}]}{{\mathcal{Z}}[0]}
\right\} \; .
\end{equation}
Indeed, in the continuum limit $\Gamma[\vec{A}^{\mathrm{ext}}]$
reduces to the vacuum energy in presence of the background field
$\vec{A}^{\mathrm{ext}}$. Moreover, the lattice background
effective action Eq.~(\ref{Gamma}) is invariant for gauge
transformations of the external links $U^{\mathrm{ext}}_k$. 
Different approaches to put background fields on the lattice
use external currents~\cite{Damgaard:1988ec,Cea:1991ag,Cea:1991td,Levi:1995kc,Ogilvie:1998my,Chernodub:2001da} 
or modified boundary conditions~\cite{Ambjorn:1989qr,Ambjorn:1990wf,Kajantie:1998rz}.

At finite temperature we are interested in  the thermal partition
function~\cite{Gross:1981br}
in presence of a static background field. On the lattice
we introduced~\cite{Cea:2001pc,Cea:2001an}
 the background field thermal partition function as
\begin{equation}
\label{ZetaTnew} \mathcal{Z}_T \left[ \vec{A}^{\text{ext}} \right]
= \int_{U_k(L_t,\vec{x})=U_k(0,\vec{x})=U^{\text{ext}}_k(\vec{x})}
\mathcal{D}U \, e^{-S_W}   \,,
\end{equation}
where the physical temperature is given by $T=1/a L_t$. On a
lattice with finite spatial extension we also impose that the
links at the spatial boundaries are fixed according to boundary
conditions Eq.~(\ref{coldwall}). Thus we see that, sending the
physical temperature to zero, the thermal functional
Eq.~(\ref{ZetaTnew}) reduces to zero-temperature Schr\"odinger
functional Eq.~(\ref{Zetalatt}).

At finite temperature the relevant quantity turns out to be the
free energy functional which is naturally defined as:
\begin{equation}
\label{freeenergy} F[\vec{A}^{\mathrm{ext}}]= - \frac{1}{L_t} \ln
\frac{\mathcal{Z}_T[\vec{A}^{\mathrm{ext}}]} {{\mathcal{Z}}_T[0]}
\,.
\end{equation}
Obviously, if the physical temperature is sent to zero the free
energy  functional reduces to the vacuum energy functional
Eq.~(\ref{Gamma}).

In our previous studies~\cite{Cea:2001an} we found that for U(1)
the confining vacuum behaves as a coherent condensate of Dirac
magnetic monopoles. On the other hand in SU(2) and SU(3) it seems
that there is condensation of Abelian magnetic monopoles and
Abelian vortices. So that in SU(2) and SU(3) gauge theories one
could look at the confining vacuum as a coherent Abelian magnetic
condensate. Moreover, we also found~\cite{Cea:1999gn}
that a constant Abelian chromomagnetic field at zero temperature is
completely screened in the continuum limit, while  at finite
temperature~\cite{Cea:2001pc} it seems that the applied field is restored by
increasing the temperature. These
results strongly suggest that the confinement dynamics is
intimately related to Abelian chromomagnetic gauge configurations.
From previous considerations we are led to investigate if deconfinement
temperature depends on the strength of an applied external
constant Abelian chromomagnetic field. To address this question 
we performed lattice simulations of pure SU(3) gauge theory 
at finite temperature and compute the free energy functional 
Eq.~(\ref{freeenergy}) for a static constant Abelian chromomagnetic field.
In the continuum 
a constant Abelian chromomagnetic field is given by:
\begin{equation}
\label{field}
\vec{A}^{\mathrm{ext}}_a(\vec{x}) =
\vec{A}^{\mathrm{ext}}(\vec{x}) \delta_{a,3} \,, \quad
A^{\mathrm{ext}}_k(\vec{x}) =  \delta_{k,2} x_1 H \,.
\end{equation}
In SU(3) lattice gauge theory we get:
\begin{equation}
\label{t3links}
\begin{split}
& U^{\mathrm{ext}}_1(\vec{x}) =
U^{\mathrm{ext}}_3(\vec{x}) = {\mathbf{1}} \,,
\\
& U^{\mathrm{ext}}_2(\vec{x}) =
\begin{bmatrix}
\exp(i \frac {g H x_1} {2})  & 0 & 0 \\ 0 &  \exp(- i \frac {g H
x_1} {2}) & 0
\\ 0 & 0 & 1
\end{bmatrix}
\end{split}
\end{equation}
To be consistent with the hypertorus geometry we impose that the
magnetic field is quantized as:
\begin{equation}
\label{quant} \frac{g H}{2} = \frac{2 \pi}{L_1}
n_{\mathrm{ext}} \,, \qquad  n_{\mathrm{ext}}\,\,\,{\text{integer}}\,.
\end{equation}
It is easy to verify that the lattice links Eq.~(\ref{t3links})
give rise to a constant field strength. 
Therefore, since the free energy functional $F[\vec{A}^{\mathrm{ext}}]$
is invariant for time independent gauge transformations of
the background field $\vec{A}^{\mathrm{ext}}$, it follows that  
$F[\vec{A}^{\mathrm{ext}}]$ is proportional to spatial
volume $V=L_s^3$ and the relevant quantity is the density
$f[\vec{A}^{\mathrm{ext}}]$ of free energy
\begin{equation}
\label{free-energy} f[\vec{A}^{\mathrm{ext}}] = \frac{1}{V}
F[\vec{A}^{\mathrm{ext}}] \,.
\end{equation}
We want to compute the density of free energy on the lattice.
In order to circumvent the problem of computing a partition function which is
the exponential of an extensive quantity, we consider the
$\beta$-derivative of $f[\vec{A}^{\mathrm{ext}}]$ (at fixed
external field strength $gH$). Indeed this quantity  is easy to
evaluate numerically since it is related to the plaquette ($\Omega
= L_s^3 \times L_t$):
\begin{equation}
\label{deriv}
\begin{split}
f^{\prime}[\vec{A}^{\mathrm{ext}}]   & = \left \langle
\frac{1}{\Omega} \sum_{x,\mu < \nu}
\frac{1}{3} \,  \text{Re}\, {\text{Tr}}\, U_{\mu\nu}(x) \right\rangle_0  \\
& - \left\langle \frac{1}{\Omega} \sum_{x,\mu< \nu} \frac{1}{3} \,  \text{Re} \, {\text{Tr}} \, U_{\mu\nu}(x)
\right\rangle_{\vec{A}^{\mathrm{ext}}} \,,
\end{split}
\end{equation}
where the subscripts on the averages indicate the value of the external field.
The generic plaquette
$U_{\mu\nu}(x)=U_\mu(x)U_\nu(x+\hat{\mu})U^\dagger_\mu(x+\hat{\nu})U^\dagger_\nu(x)$
contributes to the sum in Eq.~(\ref{deriv}) if the link $U_\mu(x)$
is a "dynamical" one (i.e. it is not constrained in the functional
integration Eq.~(\ref{ZetaTnew})).  \\
Observing that $f[\vec{A}^{\mathrm{ext}}] = 0$ at $ \beta = 0$, we
may obtain  $f[\vec{A}^{\mathrm{ext}}]$ from
$f^{\prime}[\vec{A}^{\mathrm{ext}}]$ by numerical integration:
\begin{equation}
\label{trapezu1}
f[\vec{A}^{\mathrm{ext}}]  =  \int_0^\beta
f^{\prime}[\vec{A}^{\mathrm{ext}}] \,d\beta^{\prime} \; .
\end{equation}
\begin{figure}[ht!]
\begin{center}
\includegraphics[width=0.8\textwidth,clip]{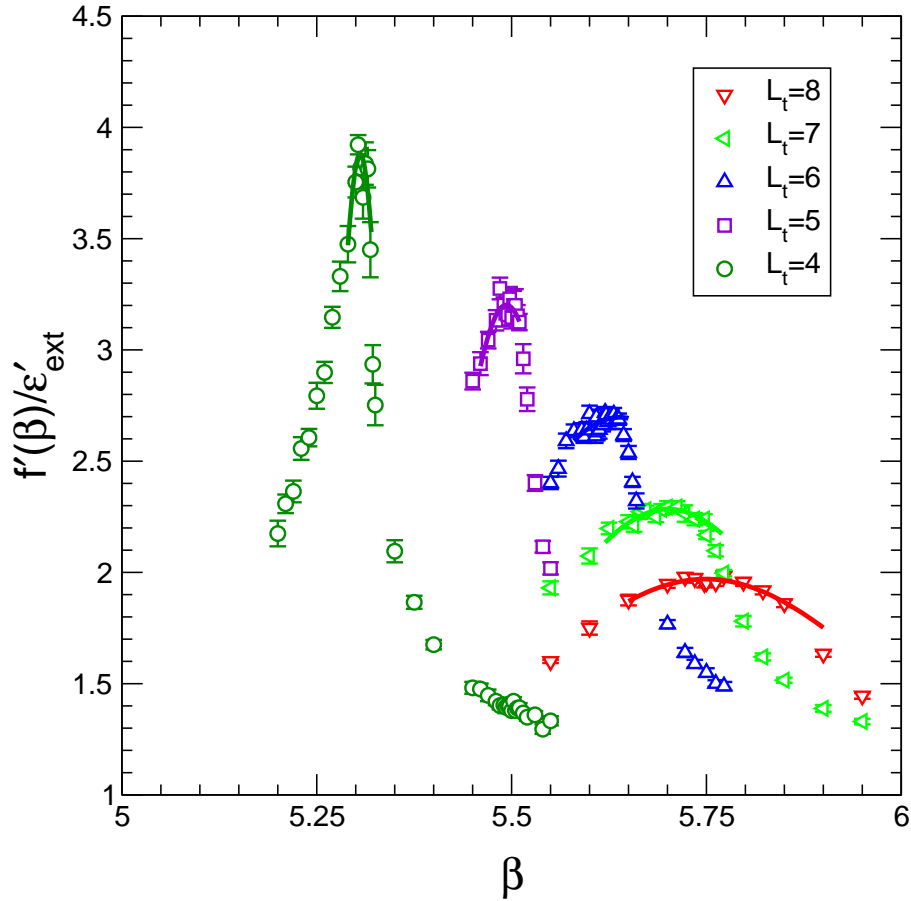}
\caption{\label{fig:01}
$f^{\prime}$ vs.  $\beta$ at fixed external field strength ($n_{\text{ext}}=1$)
for $L_t=4,5,6,7,8$ and $L_s=64$. Solid lines are the fits Eq.~(\ref{peak-form}).}
\end{center}
\end{figure}
As is well known pure SU(3) gauge system undergoes a deconfinement
phase transition by increasing temperature. As a matter of fact,
it turns out that the knowledge of
$f^{\prime}[\vec{A}^{\mathrm{ext}}]$ at finite temperature can be
used to estimate the critical temperature $T_c$. To this end it
suffices  to evaluate $f^{\prime}[\vec{A}^{\mathrm{ext}}]$ as a
function of $\beta$ for different lattice temporal sizes $L_t$. 
Indeed we found that $f^{\prime}[\vec{A}^{\mathrm{ext}}]$ displays
a peak in the critical region.
In Figure~1 we display the peak regions for different values of
$L_t$. We see clearly that the pseudocritical coupling, which is
the value of $\beta$ at the peak, depends on $L_t$. In order to
determine the pseudocritical couplings $\beta^*(L_t)$ we parameterize
$f^{\prime}(\beta,L_t)$  near the peak as:
\begin{equation}
\label{peak-form}
\frac{f^{\prime}(\beta,L_t)}{\varepsilon^{\prime}_{\mathrm{ext}}}
= \frac{a_1(L_t)}{a_2(L_t) [\beta - \beta^*(L_t)]^2 +1} \,.
\end{equation}
In Eq.~(\ref{peak-form}) we normalize $f^{\prime}$  to
$\varepsilon^{\prime}_{\mathrm{ext}}$,
the  derivative of the classical energy due to  the external
applied field
\begin{equation}
\label{epsprimeext} \varepsilon^{\prime}_{\mathrm{ext}} =
\frac{2}{3} \, [1 - \cos( \frac{g H}{2} )] = \frac{2}{3} \, [1 -
\cos( \frac{2 \pi}{L_1} n_{\mathrm{ext}})] .
\end{equation}
We restrict the region near $\beta^*(L_t)$ until the fits
Eq.~(\ref{peak-form}) give a reduced $\chi^2/{\text{ d.o.f.}} \sim
1$.
Once determined $\beta^*(L_t)$ we estimate the deconfinement
temperature as
\begin{equation}
\label{Tc} \frac{T_c}{\Lambda_{\mathrm{latt}}} = \frac{1}{L_t}
\frac{1}{f_{SU(3)}(\beta^*(L_t))} \,,
\end{equation}
where
\begin{equation}
\label{af} f_{SU(N)}(\beta) = \left( \frac{\beta}{2  N b_0}
\right)^{b_1/2b_0^2} \, \exp \left( -\beta \frac{1}{4 N b_0}
\right) \,
\end{equation}
 $N$ being the color number, $b_0=(11 N)/(48 \pi^2)$, and
$b_1=(34N^2)/(3(16\pi^2)^2)$.

In order to obtain the continuum limit critical temperature we
have to extrapolate 
$T_c/\Lambda_{\mathrm{latt}}$, given by Eq.~(\ref{Tc}),
to the continuum.
\begin{figure}[ht!]
\begin{center}
\includegraphics[width=0.8\textwidth,clip]{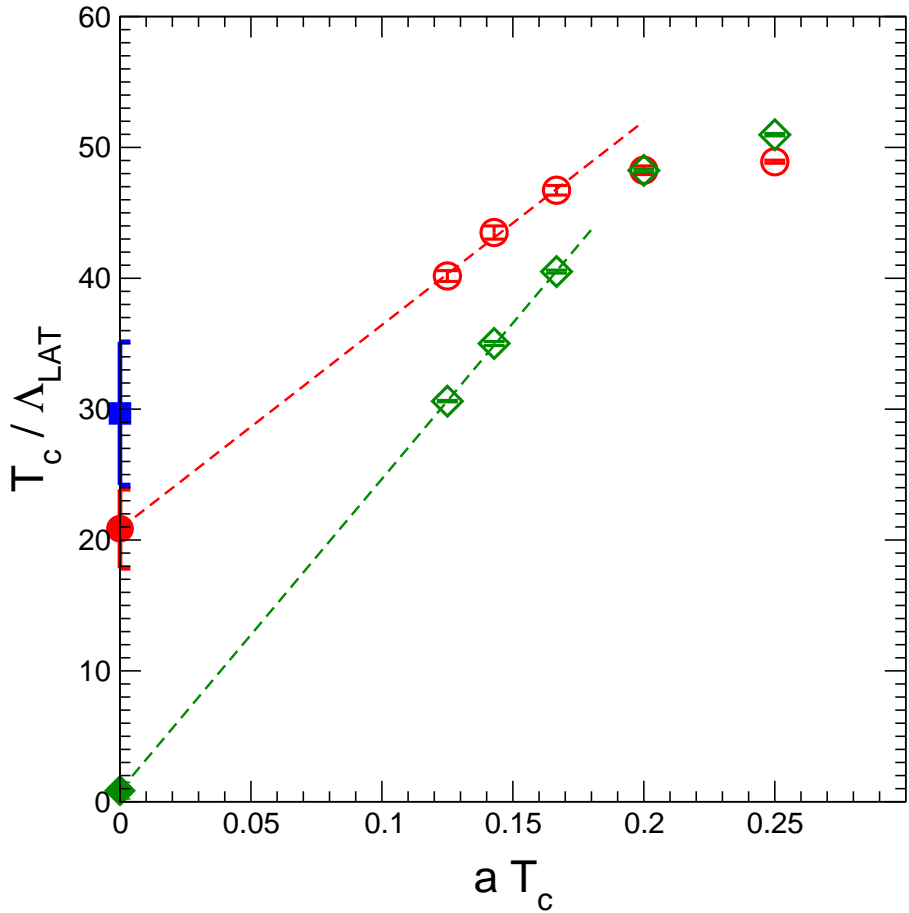}
\caption{\label{fig:02}
$T_c/\Lambda_{\mathrm{latt}}$ versus $a T_c$ for two different values
of the external field strength ($n_{\text{ext}}=1$ open circles,
$n_{\text{ext}}=10$ open diamonds). Full points refer to the continuum
extrapolations. Full square is the critical temperature Eq.~(\ref{critical}).}
\end{center}
\end{figure}
In Figure~2 we display $T_c/\Lambda_{\mathrm{latt}}$ as a function of $a T_c$
for two different values of the chromomagnetic background field. 
To extract the continuum limit of the
critical temperature we follow Ref.~\cite{Fingberg:1993ju} and perform a linear extrapolation
to the continuum of our data for $T_c/\Lambda_{\mathrm{latt}}$.
For comparison we also display  in Fig.~2 the continuum limit of the critical
temperature without background field~\cite{Fingberg:1993ju}
\begin{equation}
\label{critical}
 \frac{T_c} {\Lambda_{\mathrm{latt}}} \, = \, 29.67 \,\pm \, 5.47 \, .
\end{equation}
We see clearly that the continuum limit deconfinement critical
temperature does depend on the applied Abelian chromomagnetic
field.
Therefore we decided to vary the strength of the applied external
Abelian chromomagnetic background field to study quantitatively
the dependence of $T_c$ on $gH$~\cite{Cea:2001ef}.

To this aim we performed numerical simulations on $64^3 \times
L_t$ lattices with $n_{\text{ext}}=1,2,3,5,10$. Since we are
measuring a local quantity, such as average plaquette, a low
statistics (from 1000 up to 5000 configurations) is required in
order to get a good estimate of
$f^{\prime}[\vec{A}^{\mathrm{ext}}]$. Statistical errors have been
estimated using the jackknife resampling
method modified to take into account
the statistical correlations between lattice configurations.
As a check of possible finite volume effects we also simulated our
gauge system on $128 \times 64^2  \times L_t$ lattice with
$n_{\text{ext}}=3$. Indeed, we found that our numerical data do
not display appreciable finite volume effects.
\begin{figure}[ht!]
\begin{center}
\includegraphics[width=0.8\textwidth,clip]{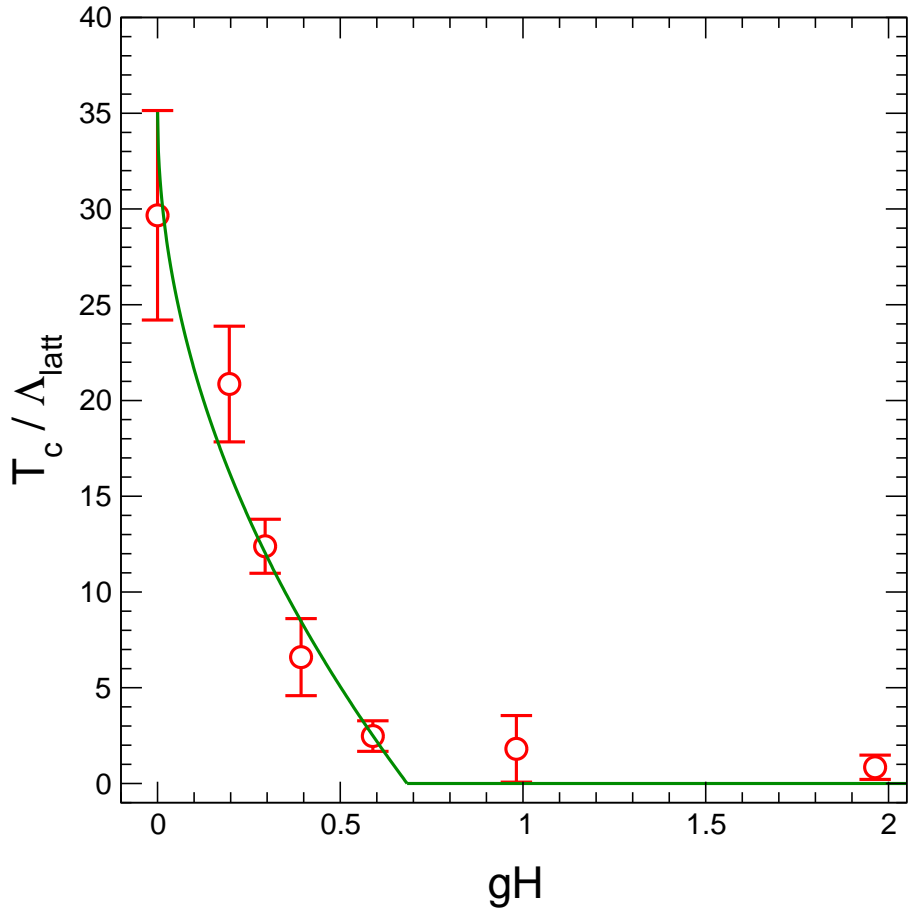}
\caption{\label{fig:03}
The continuum critical temperature $T_c/\Lambda_{\mathrm{latt}}$ versus 
the external field strength $gH$. Solid line is the fit of Eq.~(\ref{Tcfit})
to our data.}
\end{center}
\end{figure}
Following previous steps we are able to determine the critical
temperature as a function of the external chromomagnetic field. In
Figure~3 we display our determination of $T_c$ for different
values of the applied field strength. We see that the critical
temperature decreases by increasing the external Abelian
chromomagnetic field. If the magnetic length 
$a_H \sim 1/\sqrt{gH}$ is the only relevant
scale of the problem, then for dimensional reasons one expects
that
\begin{equation}
\label{dimensional-form}
T_c^2 \, \, \sim \,  \, gH \, .
\end{equation}
Indeed we try to fit our data with
\begin{equation}
\label{Tcfit} \frac{T_c(gH)}{\Lambda_{\text{latt}}} =
\frac{T_c(0)}{\Lambda_{\text{latt}}} + \alpha \sqrt{gH} \,.
\end{equation}
We find a satisfying fit (see Fig.~3) with $\alpha=-42.4 \pm 7.4$ and
$T_c(0)/\Lambda_{\text{latt}}$ that agrees within errors with
Eq.~(\ref{critical}).
Remarkably, we see that there exists a critical field
\begin{equation}
\label{Hc} gH_c \simeq 0.68
\end{equation}
such that $T_c=0$ for $gH>gH_c$.

Let us conclude by stressing the main results of this paper.
We found that there is a critical field $gH_c$ such that,
for $gH>gH_c$, the gauge system is in the deconfined phase. As a
consequence, we see  that there is an intimate connection between
Abelian chromomagnetic fields and color confinement. The existence 
of a critical chromomagnetic field is not easily
understandable  within the coherent magnetic monopole condensate
picture of the confining vacuum. On the other hand, it could be
explained if the vacuum behaves as an ordinary
relativistic color superconductor. Thus we have to reconcile two
apparently different aspects. From one hand, the confining vacuum
does display condensation of both  Abelian magnetic monopoles and
vortices, on the other hand the relation between the deconfinement
temperature and the applied Abelian chromomagnetic field
consistent with  Eq.~(\ref{dimensional-form}) suggests that the
vacuum behaves as a condensate of a color charged scalar field
whose mass is proportional to the inverse of the magnetic length.
A natural candidate for such tachyonic color charged
scalar field is the famous Nielsen-Olesen unstable
mode~\cite{Nielsen:1978rm}. The eventual condensation of the
Nielsen-Olesen modes makes the vacuum a dynamical color
superconductor. However, it should be stressed that, unlike the
case of an elementary color charged scalar field, the
chromomagnetic  condensate cannot be uniform due to gauge
invariance of the vacuum. Indeed, as showed~\cite{Feynman:1981ss} by
R.~P.~Feynman (in 2+1 dimensions)  gauge
invariance of the ground state disorders the system in such a way
that there are not long range color correlations. In this way the
disordered chromomagnetic condensate should lead to the screening
of any external Abelian chromomagnetic charge, thus explaining
the condensation of both Abelian chromomagnetic monopoles and vortices.

Another important aspect of this work might be related to
astrophysics applications. Indeed, aside from the above
considerations, our result implies that it is possible to
deconfine quarks and gluons even at low temperature and low
density. As a consequence follows the exciting possibility of
low density compact quark stars. Obviously, further progress on this
matter needs the equation of state appropriate for the description
of deconfined quarks and gluons in an Abelian chromomagnetic
field.

%\bibliography{qcd}

\end{document}